\documentclass[12pt]{iopart}

\usepackage{graphicx,bm,color,latexsym,amssymb}

\definecolor{Blue}{rgb}{0.0,0.0,1}

\definecolor{Red}{rgb}{1,0.0,0.0}

\begin{document}

\title{NMR analog of Bell's inequalities violation test}

\author{A. M. Souza$^1$}
\address{$^1$Centro Brasileiro de
Pesquisas F\'{\i}sicas, Rua Dr.Xavier Sigaud 150, Rio de Janeiro
22290-180, RJ, Brazil}
\ead{amsouza@cbpf.br} 

\author{A. Magalh\~aes$^2$, J. Teles$^2$, E. R. deAzevedo$^2$, T. J. Bonagamba$^2$}

\address{$^2$Instituto de F\'{\i}sica de S\~ao Carlos,Universidade
de S\~ao Paulo, P.O. Box 369, S\~ao Carlos 13560-970, SP, Brazil}

\author{I. S. Oliveira$^1$ and R. S. Sarthour$^1$}

\address{$^1$Centro Brasileiro de Pesquisas F\'{\i}sicas, Rua
Dr.Xavier Sigaud 150, Rio de Janeiro 22290-180, RJ, Brazil}

\begin{abstract}
In this paper we present an analog of the Bell's inequalities
violation test for $N$ qubits to be performed in a nuclear
magnetic resonance (NMR) quantum computer. This can be used to simulate or predict results for different
Bell's inequalities tests, with distinct configurations and larger number of qubits.
To demonstrate our scheme, we implemented a simulation of the violation of
Clauser, Horne, Shimony and Holt (CHSH) inequality using a two
qubit NMR system and compared the results to those of a photon
experiment. The experimental results are well described by Quantum Mechanics theory and a
Local Realistic Hidden Variables model which was specially developed for NMR. That is why we refer to  
this experiment as a {\it simulation} of the Bell's inequality violation. Our result shows explicitly how both theories 
can be compatible to each other due the detection loophole. In the last part of this work we discuss 
the possibility of testing fundamental features of quantum mechanics using NMR with 
highly polarized spins, where a strong discrepancy between quantum mechanics and 
hidden variables models can be expected. 

\end{abstract}


\maketitle

\section{Introduction}

Since the birth of Quantum Mechanics theory, interesting questions
have been raised, some of them remaining not completely
understood. One of the most amazing concerns the EPR paradox,
brought up by Einstein, Podolsky and Rosen \cite{epr}. In that
work, the authors stated that Quantum Mechanics theory is not
complete since it does not contain what they called
\textquotedblleft elements of Reality\textquotedblright . The EPR
correlations, which exists in the so-called entangled states, have
no dependence on distance, which initially led to the wrong
conclusion that they would violate the theory of relativity. One
attempt to overcome the strange features of entangled states is to
postulate the existence of some supplementary variables outside
the scope of Quantum Mechanics, called \textquotedblleft Hidden
Variables\textquotedblright \cite{genovese}. A Hidden Variables Model is supposed
to reproduce all the Quantum Mechanical predictions.

However, in 1965 John Bell \cite{bell} discovered a conflict
between Quantum Mechanics and the Hidden Variables theory.
Mathematically, this conflict takes the form of a set of
inequalities (called Bell's inequalities), which can be violated
by entangled states, but it is never violated by non-correlated
quantum states or classical ``objects''. Recently, there has been
an increasing interest in the Bell's inequalities subject, not
only to test local realism in Quantum Mechanics in a variety of
contexts, but
also because of their connection to quantum communication \cite%
{zeilinger,brassard,scarani} and quantum cryptography
\cite{acin,chen}. Furthermore, Bell's inequalities can be a useful
tool to detect entanglement, which is found to be a powerful
computational resource in quantum computation \cite{nielsen}.

Violation of Bell's inequalities has been verified in various
experiments
\cite{aspect,groblacher,wineland,monroe,rauch,yako,rachti,pan,chenpan}.
The recent development in the field of Nuclear Magnetic Resonance
Quantum Information Processing (NMR-QIP) has shown that NMR is a
valuable testing tool for the
new ideas in quantum information science (for recent reviews see \cite%
{vanderchuang,jones,coryfort,oliveira}). NMR Experiments with as many
as 12 qubits have been reported \cite{khitrin,negrevergne}. More then
fifty years of development has put NMR in an unique position to
perform complex experiments, sometimes quoted as ``spin
choreography'' \cite{freeman}. Particularly fruitful has been the use of NMR-QIP to simulate
quantum systems \cite{oliveira}.

 In this work, we use a NMR system to
simulate a quantum optics experiment. We built a scheme to
simulate the violation of Bell's inequalities for $N$ qubits
\cite{brukner,werner}, and tested it in the violation of Clauser,
Horne, Shimony and Holt (CHSH) inequality \cite{chsh} using a two
qubit NMR system. The experimental results were compared to the
Quantum Mechanical theoretical predictions and also to a Local
Realistic Hidden Variables model (LRHVM), built to explain the
correlations observed in NMR experiments \cite{menicucci}. We found 
that both theories are consistent with our experiment and that is why we refer to the experiment 
a {\it simulation}. The consistence between both theories can be understood by the fact that 
NMR can detect only a small fraction of spins due to its small polarization at room temperature. A situation that resembles 
the so called detection loophole.

It is important to stress that the NMR qubits are nuclear spins of
atoms bounded together in a single molecule, separated by few
angstroms. Therefore, a NMR experiment is inherently local and
cannot be used to prove nonlocal effects. Furthermore, most NMR-QIP
experiments are performed at room temperature in a macroscopic
liquid sample containing a large number of molecules, each of them
working as an independent \textquotedblleft quantum information
processing unit\textquotedblright . In the NMR context, the
ensemble of spins constitute a highly mixed state and their
density matrix is not entangled, as demonstrated by Braunstein et
al. \cite{braunstein}. Therefore, our work does not provide an
experimental procedure to prove or disprove nonlocal effects nor
reveal entanglement in NMR experiments at room temperature.
However, it does provide a way to simulate tests for different
Bell's inequalities. The comparison between our experiment and a
true quantum optics experiment shows the faithfully of the
simulation. Besides, our scheme can be applied to 
a highly polarized spin ensemble \cite{Anwar}. In this case true entangled states can be 
achieved and a contradiction between hidden variables models and quantum theory could be detected.  

\section{Bell's inequalities and NMR}

In this work, we refer to a generalization of the Bell's inequalities for $N$
qubits developed in \cite{brukner,werner}. It involves the
measurement of a set of correlation functions, for which, each one
of $N$ observers can choose one of the $M$ observables, whose
measurements can yield only two possibles values, $s=\pm 1$. Hence, $M^{N}$
correlation functions, named $E(n_{1},\cdots ,n_{N})$, can be
constructed, where the index $n_{i}$ runs from $n_{i}=1,\cdots
,M$, and denotes the settings of the $i^{th}$ observer. For the
NMR case, these observables are projections of the $1/2$ nuclear
spins along a particular direction labeled $n_{i}$. Taking
into account the measurement of these observables, it is possible
to build different Bell's inequalities, each of them exhibiting
contradictions with LRHVM's predictions for some entangled sates.

A general expression for the Bell's inequalities can be written as \cite%
{nagata}:

\begin{equation}
-L\leq \sum_{n_{1},\cdots ,n_{N}=1}^{M}C(n_{1},\cdots ,n_{N})E(n_{1},\cdots
,n_{N})\leq +L  \label{eq_bell}
\end{equation}
where $C(n_{1},\cdots ,n_{N})$ are real coefficients, $L$ is some limit
imposed by local realism and the correlations functions are given by:

\begin{equation}
E(n_{1},\cdots ,n_{N})=\sum_{s_{1},\cdots ,s_{N}=\pm
1}(\prod_{j}^{N}s_{j})P(s_{1},\cdots ,s_{N})  \label{cor}
\end{equation}%
being $P(s_{1},\cdots ,s_{N})$ the probability of the first
observer finding the outcome $s_{1}$, the second $s_{2}$ and so
on. In a standard experiment, a set of $N$ correlated particles is
prepared in a pure entangled sate, and their spin projection onto
$M$ different directions are measured by different observers.
After a large number of runs, the observers
compare their results in order to obtain the probabilities shown in (\ref%
{cor}) and verify whether the inequality (\ref{eq_bell}) was violated.

NMR experiments are described by density matrices of the kind
$\rho_{eq} \approx (\hat{\emph{1}} - \beta H)/2^N$, being $\beta$
the Boltzmann factor and $H$ the internal Hamiltonian of the spin
system. Only the deviation of the density matrix from unity is
observed. To use such a state to simulate the violation of
(\ref{eq_bell}) in a NMR quantum computer, the initial state is
prepared from the thermal equilibrium into a highly mixed state
called pseudo-pure state (PPS) \cite{chuang}:

\begin{equation}
\rho _{pps}=\frac{(1-\epsilon )}{2^{N}}\hat{\emph{1}}+\epsilon |\psi \rangle
\langle \psi |  \label{eq_roh}
\end{equation}%
being $\epsilon \sim 10^{-6}$, the polarization  at room
temperature. It is important to remember that the last part of
equation (\ref{eq_roh}) represents a pure state and under an
unitary transformation it behaves as such. In order to measure the spin
projection $\mathbf{r}\cdot \sigma $ onto an arbitrary direction $\mathbf{r}%
=(cos(\phi )sin(\theta ),sin(\phi )sin(\theta ),cos(\theta ))$, unitary
transformations can be used to rotate the eigenvectors of the operator $%
\mathbf{r}\cdot \sigma $ \ (being $\sigma $ a vector whose components are
the Pauli matrices $\sigma _{x}$, $\sigma _{y}$ and $\sigma _{z}$) onto the
computational basis. Since $U^{\dag }(\mathbf{r})\sigma _{z}U(\mathbf{r})=%
\mathbf{r}\cdot \sigma $ for $U(\mathbf{r})=R_{y}(-\theta )R_{z}(-\phi )$, by 
applying the appropriate $U(\mathbf{r})$ on each qubit, we have

\begin{eqnarray}
E(n_{1},\cdots ,n_{N}) &=&Tr(\rho _{pps}\mathbf{r}_{1}\cdot \sigma \otimes
\ldots \otimes \mathbf{r}_{N}\cdot \sigma )  \nonumber \\
&=&Tr(\rho^{\prime }\sigma _{z}\otimes \ldots \otimes \sigma _{z})
\label{cor2}
\end{eqnarray}%
where $\rho^{\prime }=U(\mathbf{r_{1}})\otimes \cdots \otimes U(%
\mathbf{r_{N}})\rho _{pps}U^{\dag }(\mathbf{r_{N}})\otimes \cdots
\otimes U^{\dag }(\mathbf{r_{1}})$. The above equation tells us
that the measurement of $E(n_{1},\cdots ,n_{N})$ can be achieved by
rotating each qubit by an appropriate individual rotation and then
measuring them all in the computational basis. The projective
measurement in the computational basis can be emulated by applying
a magnetic field gradient \cite{cory}, which causes the
non-diagonal elements of the density matrix to vanish. The density
matrix then becomes:

\begin{equation}
\rho ^{\prime }=\frac{(1-\epsilon )}{2^{N}}\hat{\emph{1}}+\epsilon \left[
\begin{array}{ccc}
P_{0\cdots 0} & \cdots & 0 \\
\vdots & \ddots & \vdots \\
0 & \cdots & P_{1\cdots 1}%
\end{array}%
\right]  \label{eq_diag}
\end{equation}

\begin{figure}[t]
\begin{center}
\includegraphics[width=10.0cm]{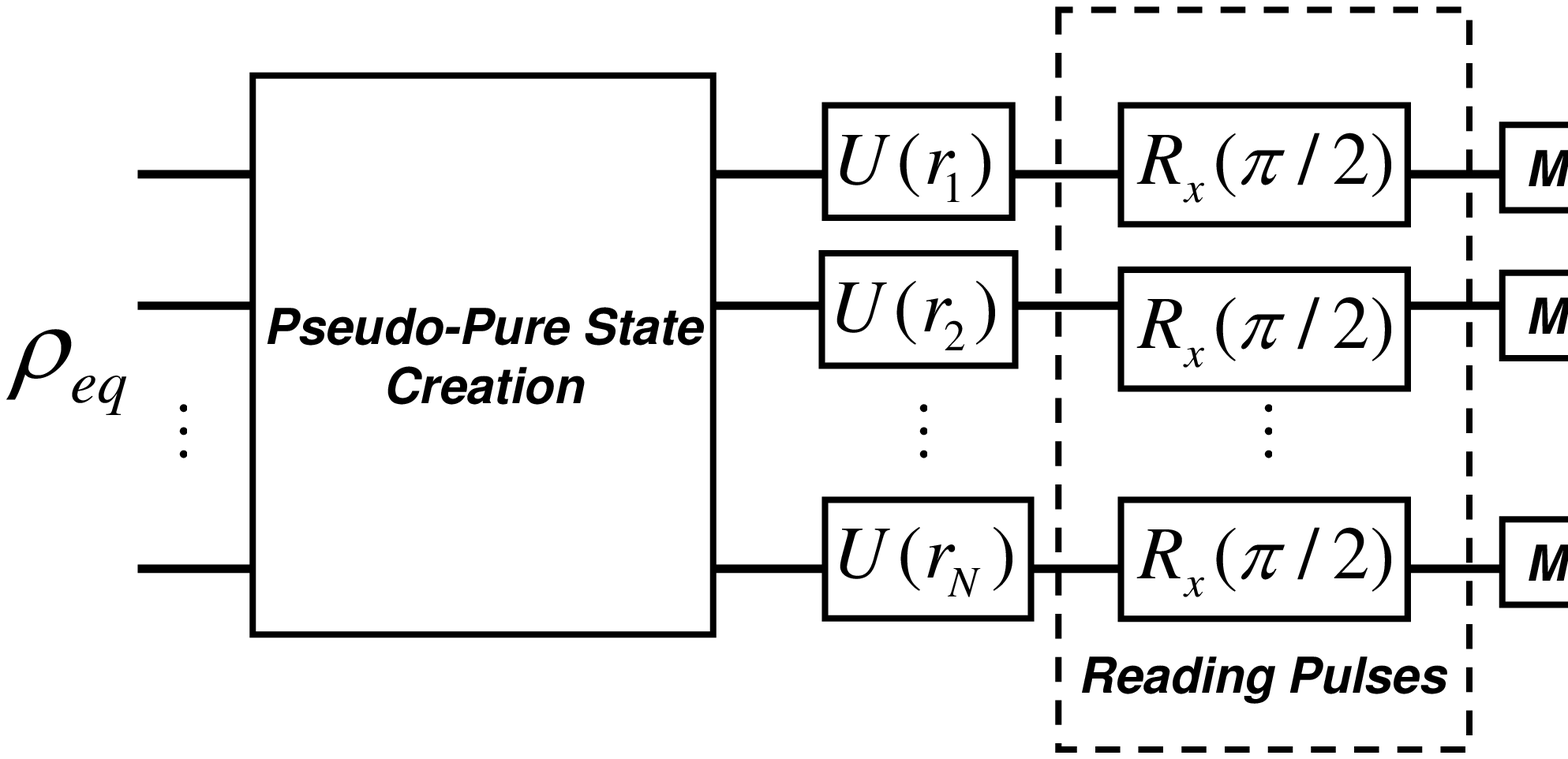}
\end{center}
\caption{ Quantum circuit proposed to simulate the correlation function of Eq. (%
\protect\ref{cor}).} \label{circ}
\end{figure}

\begin{figure}[t]
\begin{center}
\includegraphics[width=10.5cm]{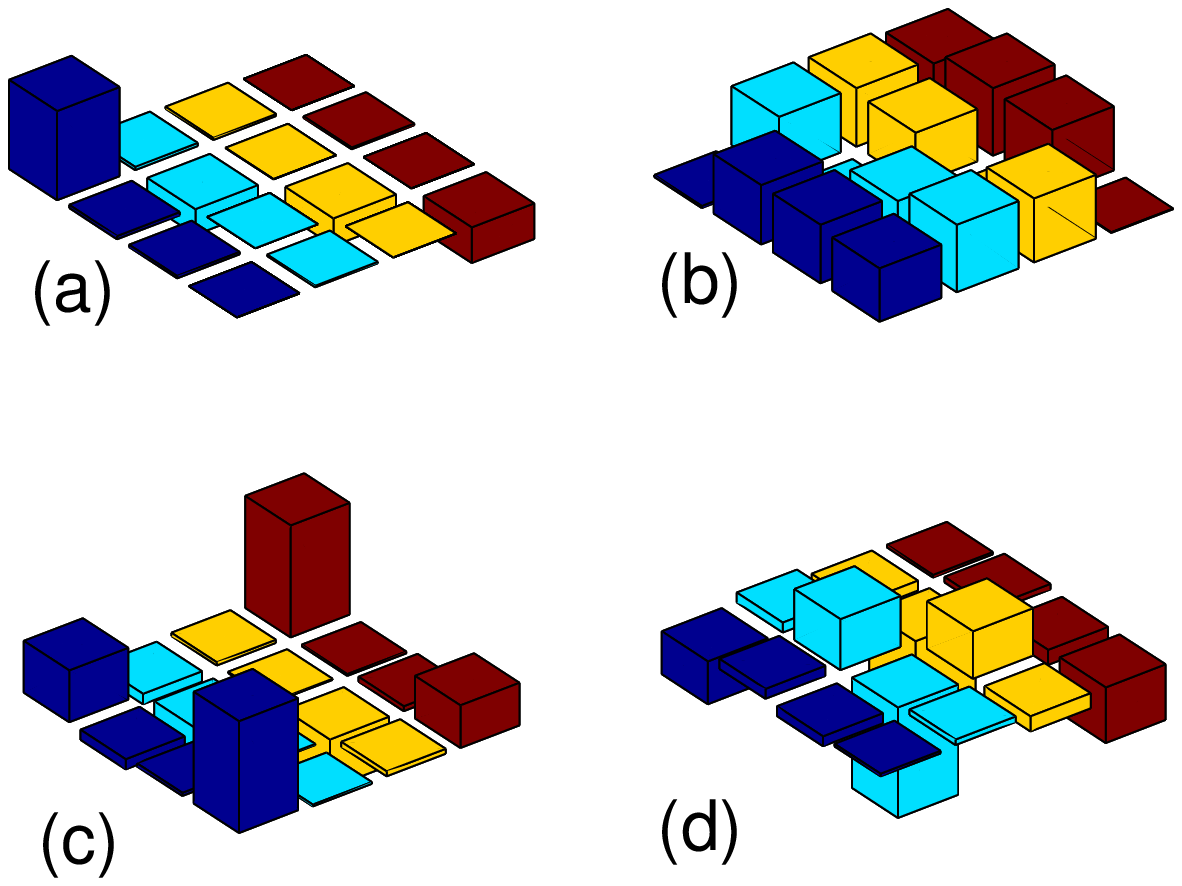}
\end{center}
\caption{The real part of deviation density matrices determined
experimentally for the investigated pseudo-pure states (a)
$\left\vert 00\right\rangle $, (b) $(\left\vert 00\right\rangle
+\left\vert
01\right\rangle +\left\vert 10\right\rangle +\left\vert 11\right\rangle)/2 $%
, (c) $(\left\vert 00\right\rangle +\left\vert 11\right\rangle)/\protect%
\sqrt{2}$, and (d) $(\left\vert 01\right\rangle -\left\vert 10\right\rangle)/%
\protect\sqrt{2} $.} \label{matrix}
\end{figure}

The populations in the second term of (\ref{eq_diag}) represent
the probabilities of finding the rotated system in one of the
$2^{N}$ energy levels. Furthermore, they are also the
probabilities $P(s_{1},\cdots ,s_{N})$ shown in (\ref{cor}), which
can be recovered from the NMR signal after applying reading pulses
to each spin. The signal detected is the average magnetization of
the sample over
time, which is proportional to the difference of populations \cite%
{freeman,oliveira}. The acquired signal is then Fourier transformed and
normalized by a reference input state. Such a normalization allows
the comparison between the experiment and theoretical results. The
scheme to measure correlation functions is shown in Figure
(\ref{circ}). The circuit must be run to each correlation function
appearing in Equation (\ref{eq_bell}).

In order to demonstrate our scheme, we used a two-qubit NMR
system, namely the nuclear spins of $^{1}$H and $^{13}$C in
chloroform (CHCl$_3$), to simulate the violation of the CHSH's
inequality \cite{chsh}, which is a special case of
(\ref{eq_bell}). It involves the measurement of the quantity

\begin{equation}
CHSH=E(n_{1},n_{2})+E(n_{3},n_{2})+E(n_{3},n_{4})-E(n_{1},n_{4})
\label{eq_chsh}
\end{equation}
where $CHSH$ is bounded by $-2\leq CHSH\leq +2$ for any LRHVM,
whereas the limits imposed by Quantum Mechanics are given by
Tsirelson's bounds $\pm 2\sqrt{2}$ \cite{tsirelson}. A particularly interesting
situation occurs when the parameters $n_{1}$, $%
n_{2}$, $n_{3}$ and $n_{4}$ labels a measurement in the directions $(0,0,1)$%
, $(sin(2\theta ),0,cos(2\theta ))$, $(sin(4\theta ),0,cos(4\theta ))$ and $%
(sin(6\theta ),0,cos(6\theta ))$, respectively. In this case,
Quantum Mechanics predicts that $CHSH=3cos(2\theta )-cos(6\theta
)$ for the pure entangled state $|\psi \rangle =(|00\rangle
+|11\rangle )/\sqrt{2}$ (the also called cat state), which results
in a maximal violation of CHSH's inequality for $\theta = 22.5^0$
and $\theta = 67.5^0$.

The NMR experiment was implemented in a Bruker Avance 500 MHz
spectrometer in the Bruker BioSpin facility in Germany. The sample contained $99 \%$ $^{13}C$ labeled
chloroform dissolved in deuterated dichlorometane
($CD_{2}Cl_{2}$), and the concentration was close to 200 mg of
$CHCl_{3}$ per 1 ml of $CD_{2}Cl_{2}$. Pseudo-pure states were
prepared by the spatial average technique, for which pulse
sequences can be found in \cite{cory}. Density matrices were
reconstructed by using quantum state tomography \cite{lee,long}. The
real parts of the experimental deviation density matrices
$\rho_{exp}$ of the investigated states are shown on Figure
(\ref{matrix}). The deviation $\delta = \frac{||\rho_{exp} - \rho_{id}||_2}{%
||\rho_{id}||_2}$ from the ideal pseudo-pure density matrices
$\rho_{id}$ are below $10\%$ in all cases, and the imaginary parts
were found to be negligible compared to the real part. The errors are mainly due to 
radio frequency field inhomogeneity and smalls pulse imperfections. The decoherence is not a important 
source of errors since the time required of entire experiment ($\sim 15$ ms) is much smaller then 
the estimated decoherence time, $T1 \sim 5$ s ($T1 \sim 15$ s)  and  $T2 \sim 200$ ms 
($T2 \sim 300$ ms) for hydrogen (Carbon).

\begin{figure}[b]
\begin{center}
\includegraphics[width=10.5cm]{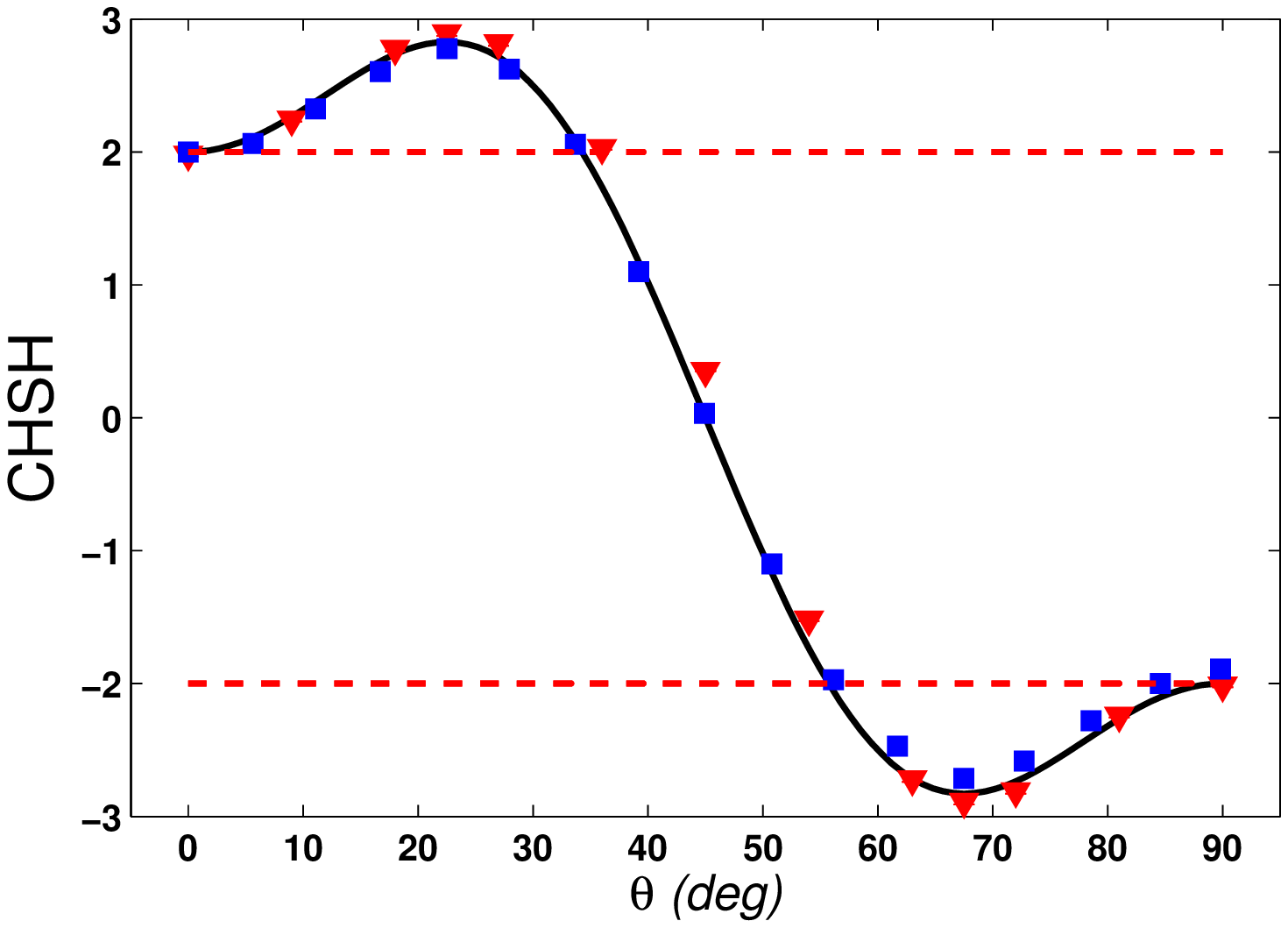}
\end{center}
\caption{Experimental results for the cat state. {\color{Red}{$%
\blacktriangledown$}} NMR experiment,
{\color{Blue}{$\blacksquare$}} Photon experiment taken from
\protect\cite{aspect}. The solid line is the Quantum Mechanical
predictions.} \label{nmrphoton}
\end{figure}

The experimental results for the cat state can be seen in Figure
(\ref{nmrphoton}), where the $CHSH$ quantity is shown as functions
of the angle $\theta $. The experimental results are also compared
to the Quantum Mechanical predictions for a pure cat state and
with a photon experiment of CHSH's inequality test extracted from
Ref. \cite{aspect}. As it can be seen, our experiment is in good
agreement with both, the Quantum Mechanics theory and the photons
experiment.

\section{Comparison with a hidden variable model}

The density matrix (\ref{eq_roh}) can be decomposed in an ensemble in which 
the fraction $\epsilon$ of the system
is in a pure state $|\psi\rangle$ while the rest
are in a completely mixed state, however it is not the unique decomposition allowed. Braunstein et al. \cite{braunstein} 
has demonstrated that any matrix of the form (\ref{eq_roh}) can be decomposed 
in a separable ensemble whenever $\epsilon \leq  1/(1+2^{N-1})$. 
This remarkable result shows that although the pseudo-pure state (\ref{eq_roh}) can 
be used to implement any quantum computation, it is classical correlated and may have a 
local realistic description which was later given explicitly in \cite{menicucci}.

In this section, we have compared our results to an explicit LRHVM
\cite{menicucci}. This model is constructed to predict the Quantum
Mechanical expectation values of any bulk-ensemble NMR experiments
that access only separable states. The most general type of
transformation of quantum states (unitary or not) can be described
via operator sum
representation $\rho \rightarrow \sum_{k}E_{k}\rho E_{k}^{\dag }$ ($%
\sum_{k}E_{k}^{\dag }E_{k}=\hat{\emph{1}}$) \cite{nielsen}. With this
formalism it is possible to simulate every step of our experiment, taking
the elements of operation $E_{k}$ as model's parameter. Starting from the
NMR equilibrium density matrix $\rho _{eq}$, we simulated every spectra and
analyzed them in the same way as we did for experimental data. The
relaxation effects were taken into account using the elements of operation
described in \cite{vandersypen}.

\begin{figure}[tb]
\begin{center}
\includegraphics[width=18.0cm]{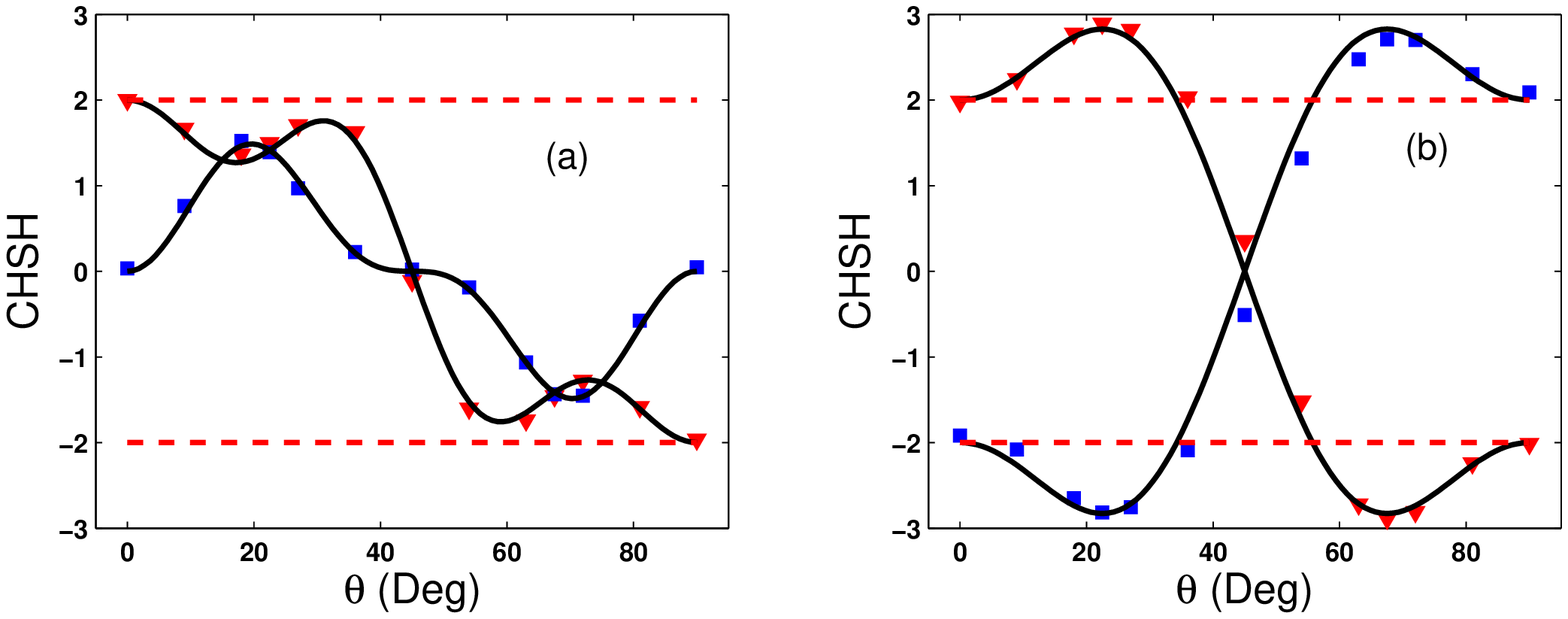}
\end{center}
\caption{Experimental results of the $CHSH$ quantity as function
of the angle $\protect\theta $. (a)
{\color{Red}{$\blacktriangledown$}} $\left\vert 00\right\rangle $,
{\color{Blue}{$\blacksquare$}} $(\left\vert 00\right\rangle
+\left\vert 01\right\rangle +\left\vert 10\right\rangle
+\left\vert 11\right\rangle)/2$. (b)
{\color{Red}{$\blacktriangledown$}} $(\left\vert (00\right\rangle
+ \left\vert 11\right\rangle)/\protect\sqrt{2} $, {\
\color{Blue}{$\blacksquare$}} $(\left\vert 01\right\rangle -
\left\vert 10\right\rangle)/\protect\sqrt{2} $. The continuous
lines are the predictions of the LRHVM described in
\protect\cite{menicucci}. The NMR data showed here are the same as in figure (\ref{nmrphoton})} \label{entsep}
\end{figure}

In Figure (\ref{entsep}a), it is shown the experimental results for 
the states $\left\vert 00\right\rangle $ and $%
(\left\vert 00\right\rangle +\left\vert 01\right\rangle +\left\vert
10\right\rangle +\left\vert 11\right\rangle)/2 $ compared with predictions
of the LRHVM described in \cite{menicucci}. As it can be seen, there is no
violation of the CHSH's inequality for these two states, as predicted by
both Quantum Mechanics theory and LRHVM, since they are separable states.

The experimental results for the states $(\left\vert
00\right\rangle +\left\vert 11\right\rangle)/\sqrt{2} $ and
$(\left\vert 01\right\rangle - \left\vert
10\right\rangle)/\sqrt{2} $ are shown in Figure (\ref{entsep}b).
Here we found a violation of the CHSH's inequality in good
agreement with Quantum Mechanics theory. Additionally, our results
are also in good agreement with the LRHVM. The fact that our experimental 
data is compatible to both theories may appear puzzling. However, it can be understood, 
noting that NMR is only sensible to the deviation part of (\ref{eq_roh}), that behaves like a 
\textquotedblleft pure entangled state" although the total ensemble is classically 
correlated as demonstrated in \cite{braunstein,menicucci}. 

This situation resembles the detection loophole , usually discussed 
in the context of optics. Generally in experiments testing Bell's inequalities 
\footnote{Up to now there is only one experiment \cite{wineland} reporting violation of Bell's 
inequality without the detection loophole.}, imperfections on the 
experimental apparatus lead to the fact that only a small sub ensemble of the total number of produced
entangled particles is actually detected. The question to be asked is whether the measured events 
is a faithful representation of the whole system. In principle, the detected sub ensemble could 
contain a distribution of hidden variables different from the total ensemble. 
Thus it is possible for the detected sub ensemble to violate the Bell's inequalities, even if the total ensemble 
do not, one can state that the sub ensemble 
\textquotedblleft simulates" the violation of Bell's inequalities. This problem, first noted in \cite{pearle}, is called the 
detection loophole. Generally, to overcome the problem,
it is invoked the fair sampling hypothesis, that state that the detected sub ensemble 
indeed represents the whole system.

In the case of NMR experiments, we cannot invoke this hypothesis, since the non-detected spins 
are known to be in highly mixed state and not in the desired entangled state. Furthermore, NMR-QIP has a known LRHVM which 
is in well agreement with experimental observation, as shown in (\ref{entsep}b). Thus our experiment is indeed a simulation.

\section{Conclusion}

In summary, we have successfully simulated a violation of a Bell's inequality
test using classical means. The faithfully of our simulation 
was tested by comparing our results with those of a photon 
experiment. We also show that we can produce the exact same set 
of data by using a LRHVM and Quantum Mechanics. This result can be viewed as a experimental demonstration on how 
both theories can be compatible due to the detection loophole. We must emphasize that such a 
LRHVM is valid only for NMR experiment, and not to photon one, although 
both curves are coincident. Besides, our protocol can be used to simulate or predict results for different
Bell's inequalities tests, with distinct configurations and larger number of qubits.

It is important to mention that the same experiment carried out in a highly 
polarized spin ensemble would not present the same features. Recently, an almost pure NMR quantum entangled state 
was achieved with polarization $\epsilon = 0.916 \pm 0.019$ \cite{Anwar}. The reported entanglement of formation of such state was  $0.822 \pm 0.039$, in this situation a true violation of Bell's inequalities is expected. Particularly interesting for NMR are those inequalities which 
do not require entanglement \cite{zela} \footnote{Note that even in this case the polarization must be enhanced because the LRHVM described in \cite{menicucci} also rules out the violation of such inequalites for $\epsilon \leq  1/(1+2^{N-1})$.}, such as the temporal Bell's inequalities \cite{leggett}, which recent proposals based on weak measurements \cite{ruskov,jordan} could be adapted to NMR systems, and those inequalities that do not require a space-like separation between the entangled particles, such as recently done in \cite{groblacher}. These inequalities are designed for the purpose to test realism.  

\begin{figure}[h]
\begin{center}
\includegraphics[width=12.0cm]{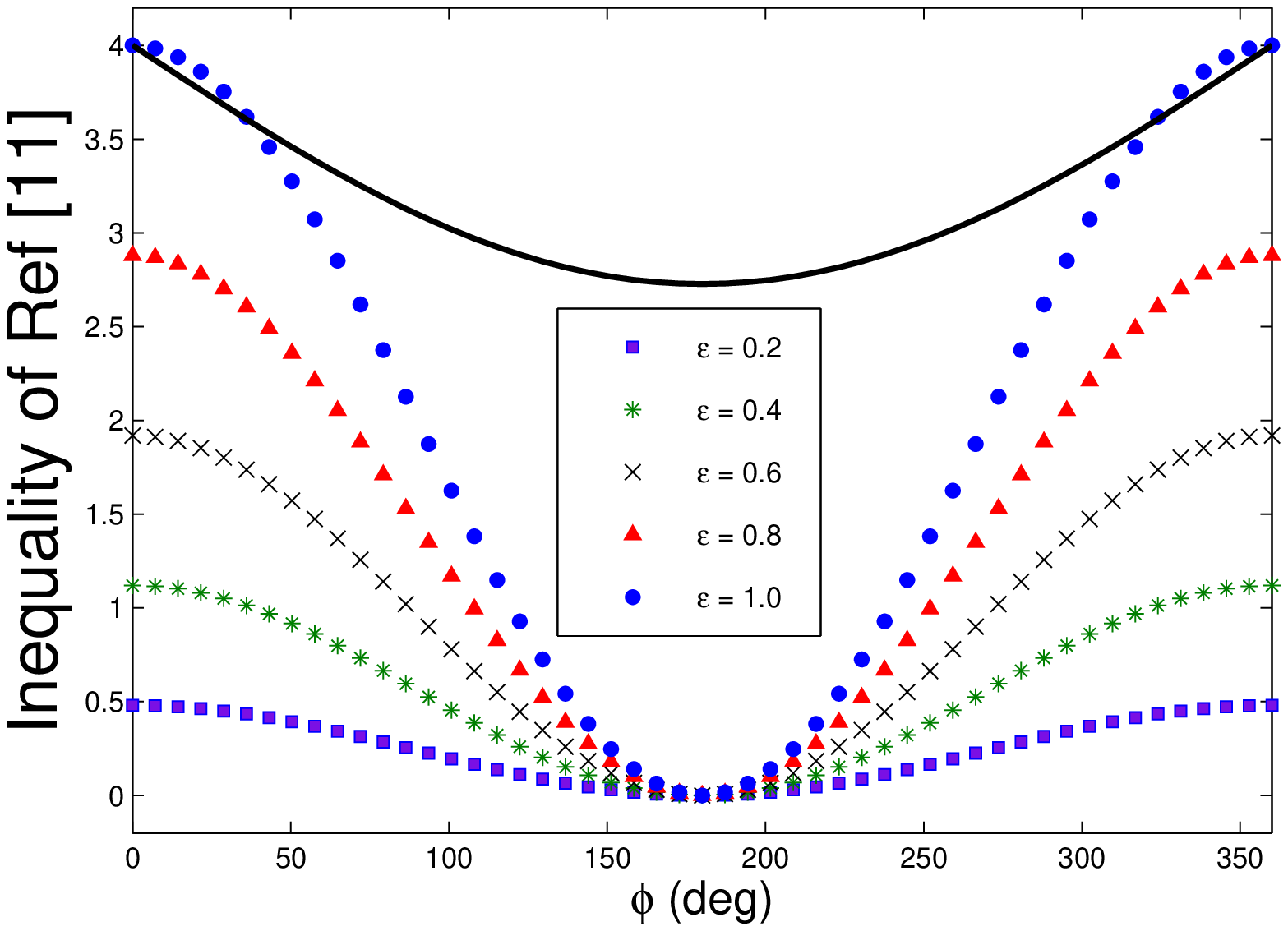}
\end{center}
\caption{Computer simulation of the violation of the inequality proposed in \cite{groblacher} for various values of the spin polarization $\epsilon$. The solid line represents the limit imposed by hidden variables, the points above the solid line may not have a realistic description.
\label{nmrrel}}
\end{figure}

In Figure (\ref{nmrrel}), we show a computer simulation of the violation of inequality found in \cite{groblacher}. We simulated the scheme described in this paper using NMR density matrix (\ref{eq_roh}) for various values of spins polarization. The solid line represent the limit imposed by the realism, the region above the limit may not have a realistic description.

We are currently seeking ways to implement these test on a NMR system. Besides the ability to simulate quantum systems. We believe that NMR quantum computation could also be used to perform real tests of quantum mechanics fundamentals. This subject is less exploired with NMR, however the ability to generate highly spin polarized ensemble allied to the high degree of control, could put NMR in a unique position in quantum information science.

\ack
The authors acknowledge support from the Brazilian funding
agencies CNPq, CAPES and FAPESP. We also like to 
acknowledge G.V.J Silva (FFCLRP $-$ USP) , R. Weisemann (Bruker BioSpin $-$ Germany),
S. Meguellatni (Bruker do Brasil) and the NMR facilities at LNLS (Campinas $-$ Brazil). This work was performed as part of the Brazilian
Millennium Institute for Quantum Information.

\section*{References}
\bibliographystyle{unsrt}
\bibliography{souza1119}

\end{document}